\documentclass[twocolumn,showpacs,preprintnumbers,amsmath,amssymb]{revtex4}
\usepackage{graphicx}
\usepackage{dcolumn}
\usepackage{bm}
\usepackage[utf8]{inputenc}

\begin{document}

\title{Long-time tails and cage effect in driven granular fluids}

\author{Andrea Fiege}
\author{Timo Aspelmeier}
\author{Annette Zippelius}
\affiliation{Georg-August-Universität Göttingen, Institut für
  Theoretische Physik, Friedrich-Hund-Platz 1, 37077 Göttingen and\\
  Max-Planck-Institut für Dynamik und Selbstorganisation,
  Bunsenstr. 10, 37073 Göttingen}

\date{\today}
\begin{abstract}
  We study the velocity autocorrelation function (VACF) of a driven
  granular fluid in the stationary state in 3 dimensions (3d). As the
  critical volume fraction of the glass transition in the
  corresponding elastic system is approached, we observe pronounced
  cage effects in the VACF as well as a strong decrease of the
  diffusion constant depending on the inelasticity. At moderate
  densities the VACF is shown to decay algebraically in time, like
  $t^{-3/2}$, if momentum is conserved locally and like $t^{-1}$, if
  momentum is not conserved by the driving. A simple scaling argument
  supports the observed long time tails.
\end{abstract}


\maketitle

Strongly agitated granular fluids have attracted a lot of attention in
recent years~\cite{poeschel}. Most of the theoretical work which is
based on microscopic dynamics has been done for either rather dilute
or weakly inelastic sytems, generalising kinetic theory to gases of
inelastically colliding particles. The velocity autocorrelation
function~\cite{dufty1} as well as transport coefficients~\cite{brey}
have been calculated for the homogeneous cooling state, which has also
been simulated for a wide range of inelasticities~\cite{dufty2}.

Comparatively few studies have been performed on the stationary state
of granular fluids in the moderate or high density regime.  This is
surprising, given the fact that the corresponding (elastic) molecular
fluids have been studied in great detail~\cite{book:hansen-1996} and
revealed several interesting features already in the dynamics of a
single tagged particle:
{\it backscattering} as indicated by a negative velocity 
autocorrelation, 
{\it long-time tails} due to the coupling of the tagged particle's
density to a shear flow and
a {\it glass transition} at a volume fraction $\eta\approx 0.58$
accompanied by a strong decrease of the diffusion constant as
a precursor to structural arrest.
It is our aim to understand which of these features pertain to an
inelastic gas and how they are destroyed by increasingly more
dissipative collisions. This applies in particular to the glass
transition, which has been conjectured to be related to the jamming
transition in granular matter~\cite{liu-nagel98}.

Several experimental groups have measured the VACF in dense granular
flow~\cite{menon, wildmann,
  PhysRevLett.92.174301,orpe:238001,urbach05,reis07}.
The VACF in the steady state of a $3$d
vibro-fluidized bed~\cite{wildmann} was shown to
exhibit strong backscattering effects.  In 2d vibrated
layers, high speed cameras have been used to measure the VACF. Even
though long-time tails seem to be beyond the experimental resolution,
these experiments give evidence for a nonexponential
decay~\cite{urbach05}. Caging effects have clearly been seen in
air-fluidized beds~\cite{Abate06} as well as in sheared
granular materials~\cite{Marty05}. In recent experiments~\cite{reis07}
the development of a plateau in the mean square displacement has been
observed, but may be related to crystallization as seen in
monodisperse vibrated layers~\cite{urbach05,reis07}.

\emph{Model---}We investigate a system of monodisperse
hard spheres of diameter $a$ and mass $m$. The time
evolution is governed by instantaneous inelastic two-particle collisions.
Given the relative velocity $\mathbf g := \mathbf v_1 -
\mathbf v_2$, the change of $\mathbf g$ in the direction $\mathbf n :=
(\mathbf r_1 - \mathbf r_2)/\left\vert (\mathbf r_1 - \mathbf r_2)
\right\vert$ is
\begin{equation}
  \left( \mathbf g \cdot \mathbf n \right)^\prime =
 -\varepsilon \left( \mathbf g \cdot \mathbf n \right)
\end{equation}
where primed quantities indicate postcollisional velocities and
unprimed ones refer to precollisional ones. The coefficient of normal
restitution $\varepsilon$ characterizes the strength of the
dissipation. For real systems, $\varepsilon$ is a function of $\mathbf
n$ and $\mathbf g$ \cite{poeschel1}. Here we consider a
simplified model with $\varepsilon = \mathrm{const.} \in
\left[0,1\right]$. The elastic system is characterized by
$\varepsilon=1$ and the sticky gas by $\varepsilon=0$. The
postcollisional velocities of the two colliding spheres are given by
${\mathbf v}_1^\prime = \mathbf v_1 - \bm \delta $ and $
 \mathbf v_2^\prime  =  \mathbf v_2 + \bm \delta$
with $\bm \delta = \frac{1+\varepsilon}{2} ( \mathbf n \cdot \mathbf g
) \mathbf n $. 

Due to the inelastic nature of the collisions, we have to feed energy
into the system in order to maintain a stationary state. This can be
done either by driving through the boundaries, like shearing the
system or vibrating its walls \cite{puglisi-2005}, or alternatively by
bulk driving like on an air table or the experiments of
ref. \cite{schroeter}. Here we choose the simplest bulk
driving~\cite{PhysRevE.54.R9} and kick a given particle, say particle
$i$, instantaneously at time $t$ according to
\begin{equation}
 \mathbf v_i^\prime (t) = \mathbf v_i(t) + v_\mathrm{Dr} \bm\xi_i(t).
\end{equation}
The driving amplitude $v_\mathrm{Dr}$ is constant and the direction
$\bm\xi_i(t)$ is chosen randomly with $\langle\xi_i^{(\alpha)}(t)
\xi_j^{(\beta)}(t^\prime)\rangle=\delta_{ij} \delta_{\alpha\beta}
\delta (t-t^\prime)$ with the cartesian components $\xi_i^{(\alpha)}$,
$\alpha=x,y,z$, distributed according to a Gaussian with zero mean. In
practice we implement the stochastic process by kicking the particles
randomly with frequency $f_\mathrm{Dr}$.

If a single particle is kicked at a particular instant, momentum is
not conserved. Due to the random direction of the kicks the time
average will restore the conservation of global momentum, but only on
average. Momentum conservation is known to be essential for the
appearance of long-time tails in elastic fluids. In fact the coupling
of the tagged particle's density to the diffusion of transverse shear
is responsible for the long-time tail of the VACF in elastic
fluids. Hence we also study a second driving mechanism, in which pairs
of particles are kicked in opposite
directions~\cite{dissip_part}. However even this kind of driving
conserves momentum only globally. To ensure momentum conservation even
on small scales, we choose pairs of neighbouring particles and kick
these in opposite directions. Our system is very close to an elastic
fluid, in the sense that we provide the thermal energy ``by hand''. We
regard this as a useful first step to investigate, which features of
an elastic molecular fluid pertain to a driven inelastic granular
fluid.

We are interested in the time delayed correlation of
a tagged particle's velocity $\Delta\mathbf v_i (t)= 
\mathbf v_i (t)-\overline{\mathbf v(t)}$ relative to the average 
$\overline{\mathbf v(t)}=1/N \sum_{i}\mathbf v_i (t)$
\begin{equation}
 \Phi(t) = \left\langle \Delta\mathbf v_i (t) \cdot\Delta \mathbf v_i(0)
 \right\rangle/
\left\langle\Delta \mathbf v_i^2 (0) \right\rangle
\end{equation}
and its mean square displacement (MSD)
\begin{equation}
  \Delta \mathbf r^2 (t) = \left\langle\left( \mathbf r_i (t)
 - \mathbf r_i(0) \right)^2 \right\rangle.
\end{equation}
Here $\langle \ldots \rangle$ denotes an average over the random noise
$\bm\xi_i(t)$. Of particular interest is the diffusion coefficient,
which is expected to decrease as we increase the volume fraction
towards close packing. It can be obtained in two alternative ways,
either via the integral of the VACF $3D= \int_0^\infty \mathrm d t
\Phi(t)$, or as the time derivative of the MSD $6D= \lim_{t\rightarrow
  \infty}\; \frac {d \Delta \mathbf r^2 (t)}{dt}$. Both definitions
are equivalent in a stationary state.

\emph{Method---}In order to determine the MSD, the VACF and the
diffusion coefficient, we performed event driven molecular dynamics
simulations for several system parameters: $0.5\leq \varepsilon \leq
0.9$ and volume fractions $0.1 \leq \eta \leq 0.53725$. To detect
long-time tails it is very important to have good statistics, since
the tails occur at times when the correlations are already quite
small. Hence we use a relatively small number of particles $N\simeq 10^4$
particles, but average each configuration over $1000$ independent
runs. Like Bizon et~al.~\cite{PhysRevE.60.4340} we choose the driving
frequency, $f_{\text{Dr}}$, of the order of the collision
frequency. The balance of energy input and dissipation
requires $v_{\text{Dr}}^2\approx \frac{1-\varepsilon^2}{4}T/m$. We choose
$v_{\text{Dr}}$ to achieve the same $T$ for different
$\varepsilon$. It is convenient to use dimensionless units such that
$a=2, m=1$ and $T=1$.
Crystallization of the system has never been observed in the
simulation, unless we prepare the system in a crystalline state
initially, which was found to be stable in time only for
$\varepsilon=0.9$ and the highest density ($\eta=0.53725$)
investigated.

Event driven simulations of dense systems with a constant coefficient
of restitution are known to undergo an inelastic collapse. Several
mechanisms have been suggested to avoid the inelastic collapse. Here
we proceed as follows: We introduce a virtual hull of very small width
for each sphere. Two approaching spheres then collide three times: When
the virtual hulls first touch each other, there is no change in
momentum; then the real spheres collide \textit{elastically} when they
touch; finally the inelastic change of momentum takes place when the
virtual hulls touch upon receding. Thus the dissipation takes place
only when the colliding particles are sufficiently separated, i.e. by
the width of the hull, 
which is taken to be $10^{-5}$ of the particles's diameter.

\begin{figure}
 \includegraphics[width=0.5\textwidth]{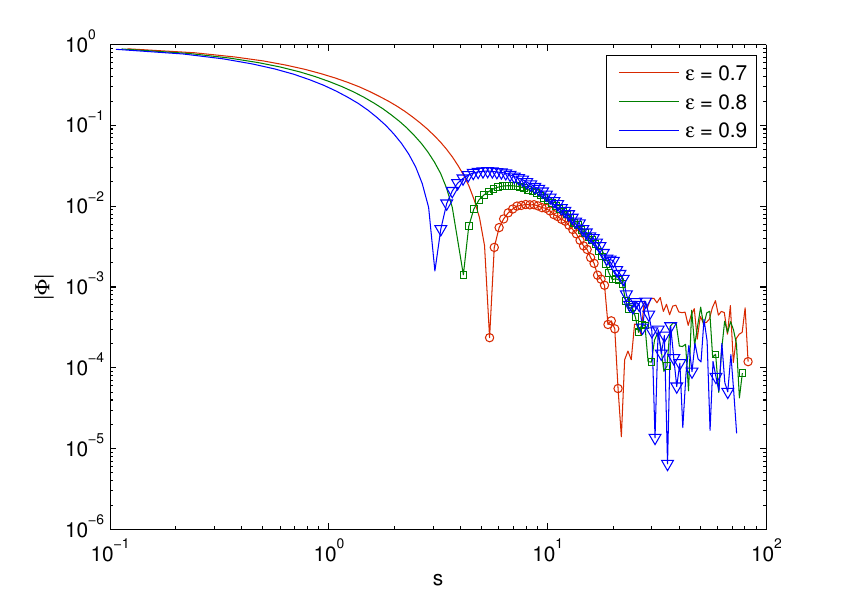}
 \caption{(color online) Modulus of the VACF for $\eta=0.50$ and
   different values for $\varepsilon$ as a function of the mean number
   of collisions per particle, denoted by $s$; symbols indicate
   negative values of the VACF.}
\label{vacf3}
\end{figure} 

\begin{figure}
 \includegraphics[width=0.5\textwidth]{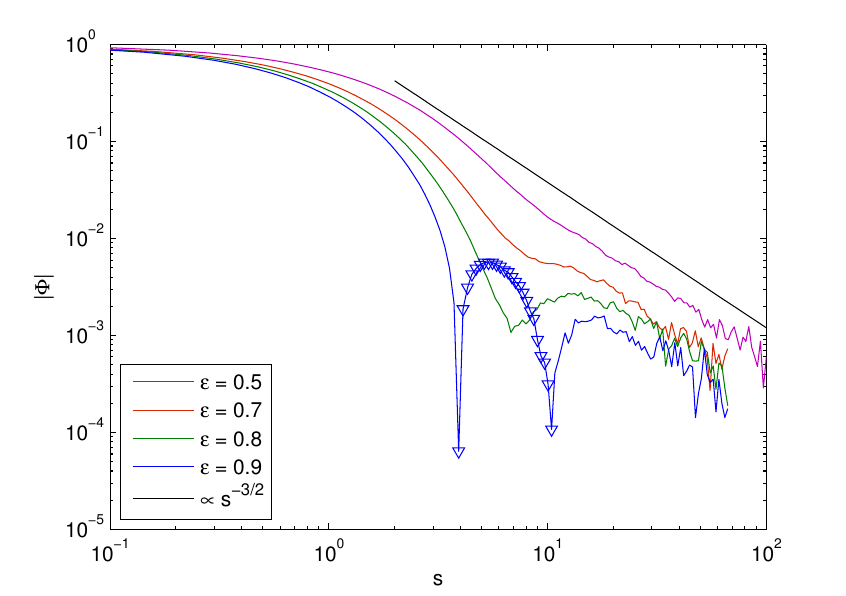}
 \caption{(color online) Modulus of the VACF for
   $\eta=0.45$ and $ 0.5 \leq \varepsilon \leq 0.9$ (from top to
   bottom); symbols in the curve for $\varepsilon=0.9$
   indicate negative values of the VACF.}
\label{vacf2}
\end{figure}

\emph{Results---}Backscattering effects are expected to be strongest
for high densities, when cages have formed locally, enforcing
reflection of the tagged particle by neighbouring particles of the
cage. In Fig.~\ref{vacf3} we show the modulus of the VACF for volume
fraction $\eta=0.5$ and different inelasticities $0.7 \leq \varepsilon \leq
0.9$. The VACF becomes negative after a few collisions for all 3 of
values of $\varepsilon$.  It stays negative for about 10 collisions
before the correlations become positive again for large times. For
increasingly more inelastic collisions the range of negative
correlations decreases and disappears completely for strongly
inelastic systems, as demonstrated in Fig.~\ref{vacf2} for
$\eta=0.45$. One clearly observes oscillations for $\varepsilon =0.9$,
wheras for $\varepsilon =0.8$ the VACF stays positive, but shows a
pronounced dint. For still smaller $\varepsilon $ backscattering
disappears completely due to two effects. First, in the sticky limit a
tagged particle is no longer reflected from its cage. Second, in order
to achieve a stationary state of the same temperature, the driving
force has to be increased for increasing inelasticity. Thereby the
system is more strongly randomized and the cages are destroyed more
frequently.

\begin{figure}
 \includegraphics[width=0.5\textwidth]{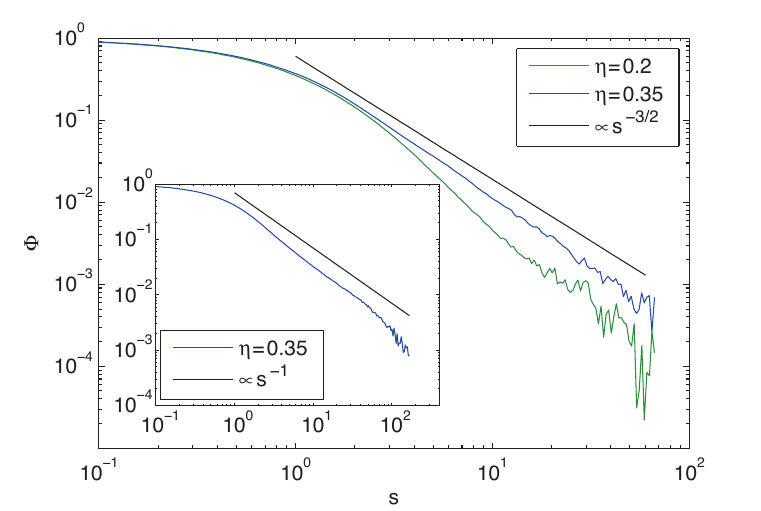}
 \caption{(color online) The VACF for a system of inelasticity
   $\varepsilon =0.7$ and volume fractions $\eta= 0.2, 0.35$ (from
   bottom to top) with local momentum conservation; inset: without
   conservation of momentum.}
\label{vacf1}
\end{figure}

Long time-tails are most easily observed for intermediate densities
($\eta=0.2,0.35$) and/or rather inelastic
systems 
such that backscattering effects do not interfere significantly.  In
Fig.~\ref{vacf1} we plot the modulus of the VACF for driving which
conserves momentum locally. For a volume fraction of $\eta=0.35$ an
algebraic tail is clearly visible and the exponent is approximately
$-3/2$ as in the molecular fluid. For volume fraction $\eta=0.2$, the
algebraic decay is shifted to larger times. To observe long time tails
for higher volume fractions one has to increase the inelasticity.  In
Fig.~\ref{vacf2} a long-time tail $\propto s^{-3/2}$ is clearly
visible only for the most inelastic system with $\varepsilon=0.5$.
In the inset of Fig.~\ref{vacf1} we plot the VACF for a
driving mechanism with random kicks of single particles, such that
momentum is not conserved. One clearly observes an algebraic decay,
however the exponent is approximately $-1$ and clearly distinct from
$-3/2$.  

A simple scaling argument~\cite{book:hansen-1996} yields the
long time tail of an elastic fluid and is easily generalised to the
driven granular system with either momentum conservation or not. Let
us assume that at $t=0$ the tagged particle has velocity $v_{i,x}$ in
the x-direction. After a short time $t$ the velocity is shared among
the $N_{t}=\rho V_{t}$ particles in a small volume $V_{t}$ around the
tagged particle: $v_{i,x}(t)\sim v_{i,x}(0)/N_{t}$. If the driving
conserves momentum locally, then the only process is diffusion of
transverse momentum, which gives rise to a diffusive growth of the
radius of $V_{t}$. This implies $V_{t}\sim t^{3/2}$ and consequently
$v_{i,x}(t)\sim t^{-3/2}$. If on the other hand, momentum is not
conserved by the driving, we have two competing processes. Collisions
among the particles still conserve momentum and give rise to the same
spread as above: $N_{t}\sim t^{3/2}$. At the same time momentum builds
up due to the random driving such that $P_{t}\sim t^{1/2}$.
Considering {\it both}, the diffusive transport due to collisions and
the built up of momentum due to driving, we find
$v_{i,x}(t)=P_{t}/N_{t}\sim t^{-1}$.

The VACF has been studied previously for a freely cooling
gas~\cite{puri07,hayakawa}, which is not in a stationary state so that
the VACF in general depends on two time arguments, the waiting time
$t_W$ and the delay $t$. For 2-dimensional systems a long-time tail
of the form $t^{-1}$ was observed~\cite{puri07}. Sheared granular
gases were found to exhibit long-time tails in the VACF with an
algebraic decay $\propto t^{-3d/2}$~\cite{kumaran:258002} in spatial
dimension $d=2$ and $d=3$.

\begin{figure}
 \includegraphics[width=0.5\textwidth]{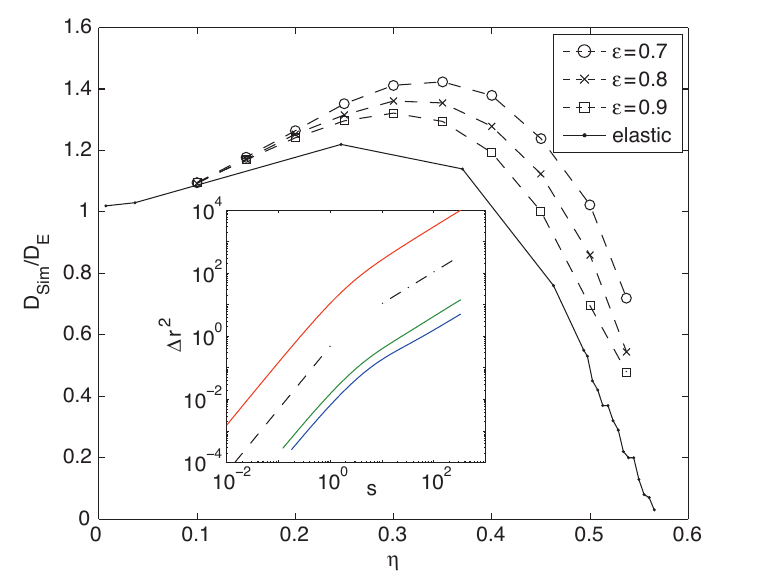}
 \caption{(color online) Diffusion coefficients relative to the Enskog
   values as a function of $\eta$; reference values for the elastic
   system from~\cite{art:speedy-1987}; inset: MSD for $\varepsilon =
   0.7$ and $\eta= 0.1$ (red), $0.5$ (green) and $0.53725$ (blue) as a
   function of $s$; the dashed lines indicate ballistic ($s^2$) and
   diffusive ($s$) behaviour.}
\label{DK}
\end{figure}

In the inset of Fig.~\ref{DK} we show a double logarithmic plot of the
mean squared displacements for a system of inelasticity $\varepsilon =
0.7$ and volume fractions $\eta= 0.1, 0.5$ and $0.53725$. The ballistic
regime can be clearly seen for up to one or two collision times. For
larger times there is a crossover to the linear regime. Even for the
dense systems no plateau is visible, although a hint of a developing
plateau may be observed for the highest density of $\eta=0.53725$. For
the longest times the MSD grows linearly with time allowing us to
extract the diffusion coefficient which decreases roughly by a factor of
20, when the volume fraction is increased from $\eta= 0.1$ to $\eta=
0.53725$.

The simplest kinetic theory for granular gases is the Enskog
approximation, which has been employed extensively for free cooling
dynamics~\cite{brey}. It can be easily extended to driven systems~\cite{fiege}
yielding an exponential decay of the VACF in the stationary
state. This approximation implies for the diffusion coeffcient:
\begin{equation}
D_{\rm E}=\frac{1}{\sqrt{\pi}} \frac{3}{8} \frac{1}{na^2 g(a)} \frac{1}{\frac{1+\varepsilon}{2}} \sqrt{\frac{T}{m}}
\label{D_Enskog}
\end{equation}
Here $n$ denotes the number density and $g(a)$ the pair correlation function at
contact, which is usually approximated by the Carnahan-Starling
formula~\cite{carnahan:635}.
We expect to observe deviations from the Enskog theory. To
quantify these, we plot in Fig.~\ref{DK} the diffusion coefficient
$D_{\text{Sim}}/D_{\text{E}}$ relative to the Enskog value as a
function of volume fraction together with reference values for an
elastic system~\cite{art:speedy-1987}.

As in the elastic case, the dependence of $D_{\text{Sim}}/D_{\rm E}$
on the volume fraction is not monotonic, but the maximum is shifted to
higher volume fractions as compared to the elastic case. The increase
over the Enskog value for intermediate densities is stronger while the
decrease over the Enskog value at high densities is smaller as
compared to the elastic case. Nevertheless we see a pronounced
decrease of the diffusion constant as the density of the glass
transition in the elastic system is approached.


\emph{Conclusion---}We have investigated the dynamics of a tagged
particle in a granular fluid, driven to a stationary state. Increasing
the density we observe a strong decrease of the diffusion constant as
the glass transition in the elastic system is approached. Cage effects
are clearly visible at these high densities in the VACF, which was
shown to oscillate as a function of time. As expected backscattering
becomes weaker as the fluid is made more inelastic.
We have shown that long-time tails not only exist in the inelastic
fluid, but depend on whether or not the driving mechanism conserves
momentum locally. If momentum is conserved locally then momentum
transport is diffusive and the decay of the VACF is identical to the
molecular fluid, like $t^{-3/2}$. If on the other hand momentum can
build up locally, then the decay of the VACF is slowed down as
compared to the conserved case, giving rise to a $t^{-1}$ decay of the
VACF.

\begin{acknowledgments}
We thank Till Kranz, Matthias Sperl and Katharina Vollmayr-Lee for many
interesting discussions.
\end{acknowledgments}


\begin{thebibliography}{}
\bibitem{poeschel}
{\it For a recent review see e.g.}
N. V. Brilliantov and T. Pöschel, "Kinetic Theory of Granular Gases",
	Oxford University Press, Oxford, 2004;

\bibitem{dufty1} J. W. Dufty, J. J. Brey and J. Lutsko, Phys. Rev. E
  {\bf 65}, 051303, 2002;

\bibitem{brey} J. J. Brey, M. J. Ruiz-Montero, D. Cubero and
  R. Garcia-Rojo, Phys. Fluids {\bf 12}, 876, 2000; R. Garcia-Rojo,
  S. Luding and J. J. Brey, Phys. Rev. E
  {\bf 74}, 061305, 2006;

\bibitem{dufty2} J. Lutsko, J. J. Brey and J. W. Dufty, Phys. Rev. E
  {\bf 65}, 051304, 2002;

\bibitem{book:hansen-1996}
{\it For a review see e.g.}
J.-P. Hansen and I. R. McDonald, "Theory of simple liquids", Academic
Press, London, 1996;

\bibitem{liu-nagel98} A. J. Liu and S. R. Nagel, Nature {\bf 396}, 21, 1998;

\bibitem{menon}
N. Menon and D. J. Durian,
Science {\bf 275}, 1920, 1997;

\bibitem{wildmann} R. D. Wildman, J. M. Huntley, and J.-P. Hansen,
  Phys. Rev. E {\bf 60}, 7066, 1999; R. D. Wildman, J.-P. Hansen and
  D. J. Parker, Phys. Fluids {\bf 14}, 232, 2002;

\bibitem{PhysRevLett.92.174301} J. Choi, A. Kudrolli, R. R. Rosales
  and M. Z. Bazant, Phys. Rev. Lett. {\bf 92}, 174301, 2004;

\bibitem{orpe:238001} A. V. Orpe and A. Kudrolli,
  Phys. Rev. Lett. {\bf 98}, 238001, 2007;

\bibitem{urbach05} P. Melby et al., J. Phys. C {\bf 17}, S2698, 2005;


\bibitem{reis07} P. M. Reis, R. A. Ingale and M. D. Shattuck,
  Phys. Rev. Lett. {\bf 98}, 188301, 2007;

\bibitem{Abate06} A. R. Abate and D. J. Durian,
  Phys. Rev. E {\bf 74}, 031308, 2006; 

\bibitem{Marty05} G. Marty and O. Dauchot, Phys. Rev. Lett. {\bf 94},
  015701, 2005; O. Dauchot, G. Marty and G. Biroli,
  Phys. Rev. Lett. {\bf 95}, 265701, 2005;

\bibitem{poeschel1} N. V. Brilliantov and T. Pöschel, Phys. Rev. E,
  {\bf 61}, 1716, 2000; N. V. Brilliantov and T. Pöschel, Chaos {\bf 15},
  026108, 2005;

\bibitem{puglisi-2005} A. Puglisi, F. Cecconi and A. Vulpiani,
  J. Phys. Cond. Matt., {\bf 17}, S2715, 2005;

\bibitem{schroeter} M. Schroeter, D. I. Goldman and H. L. Swinney,
  Phys. Rev. E, {\bf 71}, 030301(R), 2005;

\bibitem{PhysRevE.54.R9} D. R. M. Williams and F. C. MacKintosh,
  Phys. Rev. E, {\bf 54}, R9, 1996;

\bibitem{dissip_part} P. Espanol and P. Warren,
  Europhys. Lett. {\bf 30}, 191, 1995;

\bibitem{PhysRevE.60.4340} C. Bizon et al., Phys. Rev. E {\bf 60}, 4340, 1999;

\bibitem{puri07} S. R. Ahmad and S. Puri,
  Phys. Rev. E {\bf 75}, 031302, 2007;

\bibitem{hayakawa} H. Hayakawa and M. Otsuki, Phys. Rev. E {\bf 76},
  051304, 2007;

\bibitem{kumaran:258002} V. Kumaran, Phys. Rev. Lett. {\bf 96}, 258002, 2006;

\bibitem{fiege} A. Fiege, diploma thesis, University of Göttingen
  2007;

\bibitem{carnahan:635} N. F. Carnahan and K. E. Starling,
  J. Chem. Phys. {\bf 51}, 635, 1969;

\bibitem{art:speedy-1987}
R. J. Speedy, Mol. Phys. {\bf 62}, 509, 1987;






\end{thebibliography}
\end{document}